\documentclass[prx,twocolumn,amsmath,amssymb,floatfix,footinbib,bibnotes,longbibliography]{revtex4-1}
\pdfoutput=1

\usepackage[breaklinks,colorlinks=true,citecolor=blue,linkcolor=magenta]{hyperref}
\usepackage[utf8]{inputenc}
\usepackage{soul}
\usepackage{dsfont}
\usepackage{amsmath,amssymb}

\usepackage{makecell}
\usepackage[normalem]{ulem}
\usepackage[caption=false]{subfig}
\usepackage{comment}
\usepackage{graphicx}
\graphicspath{{plots/}}
\usepackage{braket}
\usepackage[usenames]{color}
\usepackage{soul}
\usepackage{float}
\usepackage{xcolor}

\begin{document}
\title{Scaling of Fock space propagator in quasiperiodic many-body localizing systems}
\author{Soumi Ghosh$^{1,4}$}%
\thanks{equally contributed}
\email{soumi.ghosh@icts.res.in}
\author{Jagannath Sutradhar$^{1,2,3}$}
\thanks{equally contributed}
\email{sutradj@biu.ac.il}
\author{Subroto Mukerjee$^1$}
\email{smukerjee@iisc.ac.in}
\author{Sumilan Banerjee$^{1}$}
\email{sumilan@iisc.ac.in}
\affiliation{$^1$Centre for Condensed Matter Theory, Department of Physics, Indian Institute of Science, Bangalore 560012, India.\\
$^2$Department of Physics, Bar-Ilan University, 52900, Ramat Gan, Israel.\\
$^3$Department of Physics, Faculty of Natural Sciences, Ariel  University,  Ariel  40700,  Israel \\
$^4$International Centre for Theoretical Sciences, Tata Institute of Fundamental Research, Bengaluru 560089, India. 
}%
\begin{abstract}
Recently many body localized systems have been treated as a hopping problem on a Fock space lattice with correlated disorder, where the many-body eigenstates exhibit multi-fractal character. The many-body propagator in Fock space has been shown to be useful for capturing this multifractality and extracting a Fock-space localization length for systems with random disorder in real space. Here we study a one-dimensional interacting system of spinless Fermions in the presence of a deterministic quasiperiodic potential using the Fock-space propagator. From the system-size scaling of the self-energy associated with the diagonal elements and the scaling of the off-diagonal elements of the propagator, we extract fractal characteristics and FS localization lengths, respectively, which behave similarly to that in the random system. We compute the sample-to-sample fluctuations of the typical self-energy and the off-diagonal propagator over different realizations of the potential and show that the fluctuations in the self-energy distinguish quasiperiodic and random systems, whereas the fluctuations of the off-diagonal elements cannot demarcate the two types of potential. 
\end{abstract}
\maketitle

\section{Introduction}\label{sec:introduction}
Localization in isolated quantum systems in the presence of disorder and interaction has received a lot of interest in recent times owing to its fascinating non-equilibrium properties~\cite{Basko2006,Gornyi2005,VoskAltman,Abanin_review.2017,Alet_review.2018,RMP_MBLDmitry_2019}. While most many-body systems in the presence of interactions thermalize~\cite{Deutsch1991,Srednicki1994,Srednicki1999}, many body localized (MBL) systems in the presence of strong enough disorder violate the eigenstate thermalization hypothesis (ETH), show area law entanglement entropy even for high energy states, and retain their memory of initial conditions for arbitrarily long times~\cite{Basko2006,Oganesyan.2007,Bauer.2013,Znidaric.2008,Huse_review.2015,Abanin_review.2017,Alet_review.2018,RMP_MBLDmitry_2019}. MBL systems have been argued to have emergent quasi-local constants of motion called the local integrals of motion (LIOMs) or l-bits that prevent thermalization~\cite{Vosk2013,LocalSerbyn.2013,PhenomenologyHuse.2014,ConstructingChandran.2015}. The existence of the MBL phase in the presence of strong disorder has been corroborated in one dimension (1D) through a mathematical proof with a weak assumption~\cite{Imbrie2016}, a large number of numerical studies~\cite{Oganesyan.2007,Pal.2010,Kjall.2014,Luitz.2015,Serbyn.2015}, e.g., using exact diagonalization (ED), as well as experiments on cold atoms and trapped ions ~\cite{schreiber2015observation,rispoli2019quantum,lukin2019probing}. Phenomenological real-space renormalization group (RSRG) approaches \cite{VoskAltman,Potter2015,Morningstar2019,Goremykina2019,Dumitrescu2019}, that can access much larger systems compared to ED, have also been developed to study the critical properties of the MBL-to-thermal transition. However, in recent years, arguments have been put forth about the instability of the MBL phase in systems with random disorder in dimension $d>1$ due to so-called \emph{avalanches} \cite{deRoeck2017} mediated via `rare regions' of weak disorder. The stability of MBL even in 1D random systems has been called into question \cite{vsuntajs2020quantum,Polkovnikov2021,Sels2022}, and evidence of long-range many-body resonances \cite{Morningstar2022}, that can potentially destroy MBL in the thermodynamic limit, at least, over a much larger range of disorder than previously deduced \cite{Luitz.2015}, has been shown. 

Starting with the pioneering experiment of Schreiber {\em et. al.}~\cite{schreiber2015observation}, strong evidence of MBL has also been found in systems with quasiperiodic potentials in 1D ~\cite{Iyer.2013,Modak.2015,Li.2015,Setiawan.2017,Modak.2018,Deng.2017,Aramthottil2021}. The quasiperiodic potential being deterministic does not allow large rare regions of weak disorder on its own. Thus, naively, the avalanche instability should be absent in quasiperiodic systems, unlike the random case. However, in the presence of interactions, Hartree-Fock shifts, either in the many-body states or in the initial condition for the non-equilibrium time evolution, can still lead to effective rare regions \cite{Agrawal2022}. Based on ED studies \cite{Khemani2017}, the MBL transitions in the random and quasiperiodic systems have been argued to be in different universality classes due to differences in `inter-sample' (different disorder realizations), and `intra-sample' (different eigenstates for the same realization) fluctuations of entanglement entropy. Phenomenological RSRG studies \cite{Zhang2018,Zhang2019}, though consistent with two different universality classes, reach very different conclusions regarding the critical properties, e.g., the critical exponent, of the quasiperiodic MBL transition and the relevance of weak random perturbations on the quasiperiodic potential. Since the ED studies are limited to small finite-size systems, and as the phenomenological RSRG rules cannot be derived \emph{ab initio} from the microscopic models, the questions on the fundamental differences between MBL phenomena in interacting random and quasiperiodic systems remain largely unresolved.

In this work, we construct a Fock-space (FS) picture, complementary to the above real-space perspective, for the quasiperiodic MBL and the transition. In particular, we ask how the differences between the random and quasiperiodic potentials in real space manifest in the FS localization properties, namely in the many-body FS propagators \cite{Sutradhar2022.FSP,Nroy2022.NEE}.
The interacting problem in real space can be viewed as a non-interacting hopping problem on the complex FS graph or lattice with correlated disorder ~\cite{Altland2017,Welsh2018.simple,Logan2019.Fock,roy2020.fock,Roy2020,Ghosh2019.correlated,Sutradhar2022.FSP}.
This has led to approximate analytical approaches to describe MBL phenomena \cite{Pietracaprina2016,Logan2019.Fock}, insights into the importance of strong FS correlations to realize MBL ~\cite{Altland2017,Welsh2018.simple,Logan2019.Fock,roy2020.fock,Roy2020,Ghosh2019.correlated}, the demonstration of multifractality of the MBL eigenstates~\cite{deluca2013,Luitz.2015,Mace2019.multifractal,roy2021.Fock,DeTomasiRare2021}, and numerical scaling theories of the MBL transition in terms of the FS inverse participation ratio (IPR) and FS propagator ~\cite{Mace2019.multifractal,roy2021.Fock,Sutradhar2022.FSP}. The FS perspective has been useful \cite{Nroy2022.NEE} to distinguish  thermal and MBL states from other types of non-ergodic states and understand the MBL proximity effect \cite{nandkishore2015many} in a quasiperiodic system with a single-particle mobility edge.   

The FS propagator can be computed using an efficient recursive technique~\cite{Sutradhar2022.FSP} that can reach systems sizes comparable to those accessible by state-of-the art ED methods \cite{polfed,Luitz.2018}.
It has been shown that the Feenberg self-energy~\cite{Logan2019.Fock} derived from the diagonal elements of the FS propagator acts as a probabilistic order parameter for the thermal-MBL transition and captures information about the multi-fractal nature of the MBL states~\cite{Sutradhar2022.FSP}. The off-diagonal elements of the FS propagator have also been studied in Ref.~\cite{Nroy2022.NEE,Greens_new} for random systems and a quasiperiodic systems with single-particle mobility edges. This has led to the identification of an FS localization length that remains finite in the MBL phase and at the MBL-thermal transition. Thus, the FS localization length  mimics the behaviour of the length scale associated with the size of the LIOMs \cite{LocalSerbyn.2013,PhenomenologyHuse.2014,ConstructingChandran.2015} namely, the local support of a LIOM in real space. 

Here we calculate the diagonal and off-diagonal elements of the FS propagator using the recursive technique for the interacting quasiperiodic one-dimensional Aubry-Andre-Harper (AAH) model \cite{AA.model}. We show that the typical values of the Feenberg self-energy and the off-diagonal elements of the FS propagator, extracted from different parts of the FS lattice and over many disorder samples, show features very similar to the random case \cite{Sutradhar2022.FSP,Greens_new}.
For example, a fractal dimension $D_s$ extracted from the Feenberg self-energy asymptotically approaches $1$ in the thermal phase, and  varies between $0$ and $1$ in the MBL phase, signifying the multifractal nature of the many body eigenstates. Based on an asymmetric finite-size scaling ansatz \cite{Mace2019.multifractal,Sutradhar2022.FSP}, we show that a \emph{non-ergodic} volume scale in the thermal phase and a correlation length in the MBL phase diverge with a Kosterlitz-Thouless-like essential and power-law singularities, respectively, approaching the MBL transition. On the other hand, the FS localization length extracted from the off-diagonal elements of the FS propagator in the MBL phase shows a variation with potential strength very similar to that of the localization length associated with LIOMs in the MBL phase. While these features are alike in the presence of quasiperiodic and random disordered potentials, the sample-to-sample fluctuations of the Feenberg self-energy seem to differentiate between random and quasiperiodic disorders. We discuss the possible connection between the fluctuation of self-energy, related to an inverse time scale, and rare-region effects, or their absence, for random and quasiperiodic disorder. We show that the inter-sample fluctuations in off-diagonal elements associated with the FS localization length cannot distinguish between random and quasiperiodic systems, and thus, presumably remains largely unaffected by these rare region effects.




\section{Model and method}\label{sec:model}
In this paper, we consider the standard microscopic model \cite{Oganesyan.2007} for MBL systems, namely interacting spinless fermions on a one-dimensional lattice with either quasiperiodic or random onsite potential
\begin{equation}\label{eq:hamiltonian}
    \hat{H}=t\sum\limits_{i=1}^L \left(\hat{c}_i^\dagger\hat{c}_{i+1}+ h.c.\right)+\sum\limits_{i=1}^L W_i \hat{n}_i +V\sum\limits_{i=1}^{L-1}\hat{n}_i\hat{n}_{i+1}
\end{equation}
where the operator $\hat{c}_i^\dagger$ creates a spinless fermion at site $i$. The model can be equivalently described in terms of hard-core bosons or as the XXZ model for spin-1/2 via Jordan-Wigner transformation.
 The quasiperiodic potential is taken to be the Aubry-Andre-Harper potential~\cite{AA.model}, $W_i=W \mathrm{cos}(2\pi \alpha i+\phi)$, where $\phi$ is a random phase chosen from the uniform distribution $\mathcal{U}(-\pi,\pi)$ and the irrational number $\alpha=(\sqrt{5}-1)/2$, the inverse of golden ratio. The hopping parameter $t$ and the nearest-neighbor interaction strength $V$ are set to $0.5$ and $1$, respectively, to be consistent with numerical studies of the XXZ spin model \cite{Luitz.2015}. For the random potential, $W_i\in [-W,W]$ follows a uniform distribution with strength $W$. Both random and quasiperiodic models show signatures of thermal (ergodic) to MBL (non-ergodic) transition in ED studies for finite systems as the potential strength $W$ is tuned \cite{Luitz.2015,Iyer.2013,Aramthottil2021}. The critical potential strengths for these transitions have been estimated as $W_c\sim 3 - 4$ and $W_c\sim 2 - 3$, for random and quasiperiodic systems, respectively. 
 \begin{figure}[h!]
     \centering
     \includegraphics[width=0.45\textwidth]{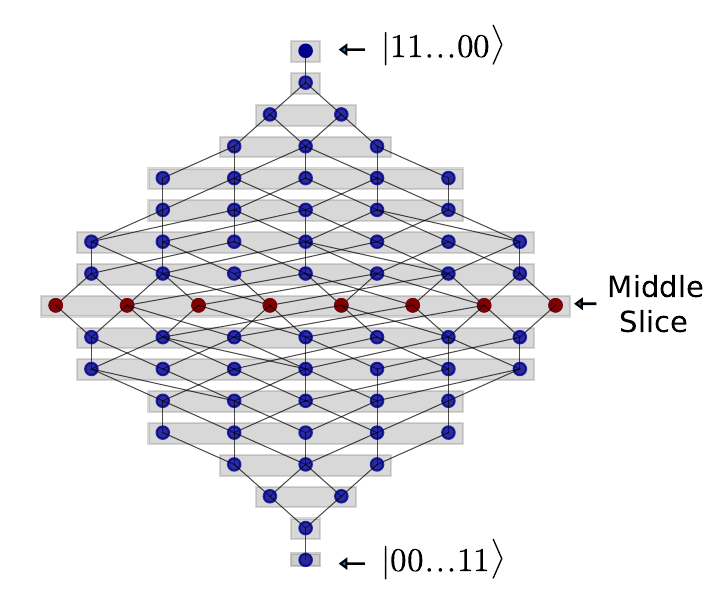}
     \caption{Schematic representation of the Fock space (FS) lattice for system size $L=8$. The blue dots represent different basis elements, and the black lines denote the hopping connections generated by the nearest-neighbor hopping in the real space one-dimensional lattice. The schematic illustrates the layered structure described in the text. The two apexes denote the minimally connected FS sites $|11110000\rangle$ and $|00001111\rangle$. The FS sites within the middle slice (largest slice) are highlighted in red.}
     \label{fig:FS-lattice}
 \end{figure}
 
 Most of the numerical studies leading to the current understanding of these systems across the thermal-MBL transition have been obtained by analyzing different properties of the many-body energy spectrum, eigenstates, and their entanglement characteristics \cite{Oganesyan.2007,Pal.2010,Bardarson2012,Serbyn2013,Kjall.2014,Serbyn.2015,Huse_review.2015,Luitz.2015}, as well as dynamical correlations \cite{Znidaric.2008,Mierzejewski2016,Bera2017}, obtained via ED. Here, we study the many-body resolvent or the Green's function defined as
\begin{equation}\label{eq:green}
    \mathbf{G}(E)=\left[E^+\mathbf{\mathbb{I}}-\mathbf{H}\right]^{-1}.
\end{equation}
Here, $E^{+}=E+\mathrm{i}\eta$, where $E$ represents the energy of interest and $\eta>0$ acts as a broadening parameter for the delta function in the calculation. We use a recursive technique~\cite{MacKinnon1983.scaling} for computing the many-body Green's function following Refs.~\cite{Sutradhar2022.FSP,Nroy2022.NEE,Greens_new}. For implementing the recursive technique, we first write the Hamiltonian in Fock space (FS), which is a connected graph representing the Hilbert space of the many-body problem~\cite{Welsh2018.simple}. Each FS site corresponds to a set of occupation numbers for all the real space sites: $|I\rangle=|n_1^{(I)},n_2^{(I)},\dots,n_L^{(I)}\rangle$. For the case of half-filling considered, there are $\mathcal{N}_{F}=\binom{L}{L/2}$ sites in the many-body Hilbert space of a one-dimensional lattice with $L$ sites. When expressed in the FS basis, the Hamiltonian [Eq.~\eqref{eq:hamiltonian}] takes the form of a tight-binding model~\cite{Welsh2018.simple,Ghosh2019.correlated,Logan2019.Fock} with correlated onsite disorder, 
\begin{equation}\label{eq:FS_H}
    H=\sum \limits_{I,J} T_{IJ}|I\rangle\langle J|+\sum\limits \mathcal{E}_I |I\rangle\langle I|,
\end{equation}
where $T_{IJ}$ indicates hopping connections between FS sites $I$ and $J$, and $\mathcal{E}_I$ denotes onsite potential. Specifically, $T_{IJ}$ is equal to $t$ if $I$ and $J$ are connected by a nearest-neighbour hop in real space [Eq.\eqref{eq:hamiltonian}], and zero otherwise. The onsite disorder potentials, $\mathcal{E}_I=\sum_{i=1}^{L} W_i n_i^{(I)}$, are highly correlated, i.e., $\left<\mathcal{E}_I\mathcal{E}_J\right>\neq 0$ for typical $I,J$ pairs~\cite{Altland2017,Ghosh2019.correlated,roy2020.fock}. These correlations are essential for localization on the FS lattice, especially given its typical coordination number diverges in the thermodynamic limit. For the Hamiltonian in Eq.\eqref{eq:hamiltonian}, Fig.~\ref{fig:FS-lattice} illustrates the layered structure of the FS, with $(1+L^2/4)$ number of slices, each containing different FS sites~\cite{Welsh2018.simple,Sutradhar2022.FSP}. In such a layered arrangement, an FS site in one slice connects only to sites in adjacent slices.
Leveraging this layered structure and the tight-binding nature of the Hamiltonian, the recursive technique, detailed in Ref.~\onlinecite{Sutradhar2022.FSP}, computes the different Green's function elements $G_{IJ}(E)=\langle I|\mathbf{G}(E)|J \rangle$ iteratively. Each iteration adds a single FS layer, facilitating computations for system sizes $L\lesssim 22-24$, comparable to the \emph{state-of-the-art} ED calculations \cite{polfed,Luitz.2018}, especially when prioritizing specific elements.


We shift the energy eigenvalues for each realization of the Hamiltonian using the transformation $H \rightarrow H-\left(\mathrm{Tr}H/\mathcal{N}_F\right)\mathbb{I}$ to keep the middle of the spectrum aligned for different disorder realizations. The many-body density of states for such a system is a Gaussian distribution with the variance $\mu_E$ scaling as $\propto L$ with increasing system sizes \cite{Logan2019.Fock}. Consequently, we rescale the Hamiltonian as $H\rightarrow H/\sqrt{L}$ to make the density of states invariant to system sizes~\cite{Logan2019.Fock,Sutradhar2022.FSP}. 

For our analysis, we calculate the diagonal elements $G_{II}$ and also several off-diagonal elements $G_{IJ}$ and $G_{1I}$ of the Green's function (Eq.~\ref{eq:green}). Here both $I$ and $J$ belong to the middle slice of the FS lattice (Fig.~\ref{fig:FS-lattice}). Thus, the particular choice of $G_{IJ}$ characterizes the non-local FS propagator between generic sites at the middle slice with connectivity $\propto L$. On the contrary, $G_{1I}$ captures the propagation between a site in the middle slice $I$ and a \emph{non-generic} FS site $1\equiv|1111\dots 0000\dots \rangle$ located at the apex of the FS graph, characterized by the minimal connectivity (exactly $1$).
Given our focus on the spectrum's midpoint, we set $E=0$.
The broadening $\eta$ is chosen as the mean many-body level spacing $\delta=\sqrt{2\pi}\mu_E/\sqrt{L}\mathcal{N}_F$. For statistical purposes, we average all the quantities over different realizations of the quasiperiodic potential by choosing different phases $\phi$. Specifically, we choose $10000$, $5000$, $2000$, $1000$, and $100$ samples for system sizes $12$, $14$, $16$, $18$, and $20$, respectively. For the random disorder discussed in Sec.~\ref{sec:results}, we consider identical system sizes and disorder realizations as in the quasiperiodic case.

\section{Results}\label{sec:results}
Here we first discuss different aspects of the diagonal and the off-diagonal elements of Green's function for the quasiperiodic potential in Eq.\eqref{eq:hamiltonian}. 
The distinct system size scalings of these diagonal and off-diagonal elements in the thermal and MBL phases are discussed in subsections~\ref{ssec:self_en} and \ref{ssec:FS_llength}. We also discuss some characteristic features of the distribution of the diagonal elements of Green's function in subsection~\ref{ssec:distribution}
Finally, in subsection~\ref{ssec:fluct}, we discuss the sample-to-sample fluctuations of these Green's function elements, drawing comparisons with their counterparts in the random disordered case.

\subsection{On-site Feenberg self-energy}\label{ssec:self_en}

We calculate the on-site self-energy, i.e., the Feenberg self-energy, \cite{Logan2019.Fock} denoted as $\Sigma_I(E)$ corresponding to site $I$ from the diagonal element of the Green's function as $G_{II}(E)=\langle I |\mathbf{G}(E)|I\rangle=\left[E+\mathrm{i}\eta -\mathcal{E}_I-\Sigma_I(E)\right]^{-1}$. We are specifically interested in the imaginary part of the self-energy $\Delta_I(E)=\mathrm{Im}\Sigma_I(E)$, henceforth referred to as the Feenberg self-energy for brevity. We calculate the typical value of the Feenberg self-energy defined as $\ln \Delta_t=\langle \ln \Delta_I\rangle_{I,\phi}$, where $\langle \cdots\rangle_{I,\phi}$ represents averaging over all FS sites belonging to the middle slice and different values of $\phi$. The self-energy $\Delta_I(E)$ represents the inverse lifetime of an \emph{excitation} of energy $E$ created at the FS site $I$ \cite{Sutradhar2022.FSP}. Therefore, in the thermal phase, we expect the excitation to decay in some finite time and consequently $\Delta_t\sim O(1)$ in the thermodynamic limit. On the other hand, in the MBL phase, where we expect the state to survive for an infinitely long time, the typical value $\Delta_t\rightarrow 0$ in the thermodynamic limit. It has been shown~\cite{Sutradhar2022.FSP,Nroy2022.NEE} that due to the multifractal nature~\cite{Mace2019.multifractal} of the MBL states, the typical value $\Delta_t$ decreases with increasing system sizes as $\Delta_t\sim \mathcal{N}_{F}^{-(1-D_s)}$ for $\eta\sim \mathcal{N}_F^{-1}\ll \eta_c$, where $\eta_c\sim \mathcal{N}_{F}^{-z}$ ($0<z<1$) is an emergent energy scale and $0<D_s<1$ is a spectral fractal dimension~\cite{Altshuler2016a}.

\begin{figure}[h!]
\includegraphics[width=0.5\textwidth]{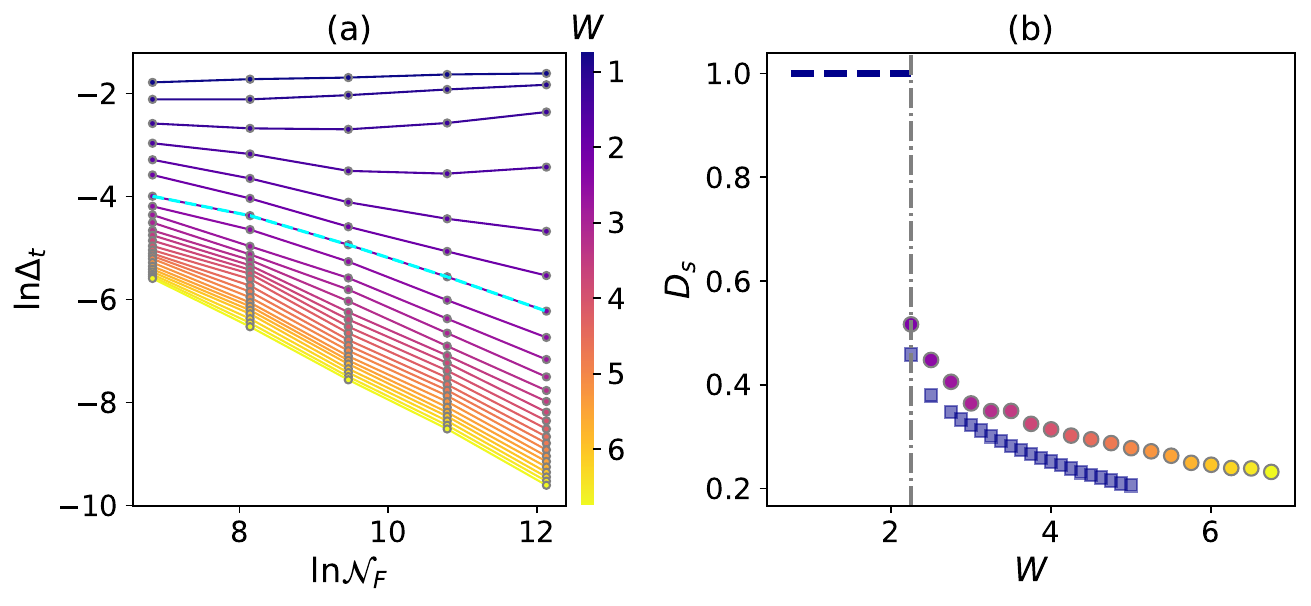}
\caption{(a) Variation of the typical value $\Delta_t$ with increasing system size. In the thermal phase ($W<W_c$), $\Delta_t\sim O(1)$, while in the MBL phase ($W>W_c$), $\Delta_t$ decreases following a power-law $\Delta_t\propto \mathcal{N}_F^{-(1-D_s)}$, with $D_s$ being the fractal dimension. The black dashed line corresponds to the critical potential strength $W_c$. (b) Variation of fractal dimension $D_s$ as a function of the potential strength where $0<D_s<1$ in the MBL phase including the critical point. $D_s$ extracted from the power-law fit of the raw data of panel (a) [circle] and from finite-size scaling [square] are both shown. The critical potential strength $W_c$ is indicated by the vertical dashed line.}
\label{appfig:fractal}
\end{figure}

The multifractal nature and the underlying scaling of the self-energy in both the thermal and MBL phases have been investigated in Ref.~\cite{Sutradhar2022.FSP} for an interacting one-dimensional system of spinless fermions in the presence of random disorder. Here, we perform a similar analysis in the presence of a quasiperiodic potential and study whether the nature of the disorder in real space, namely random vs. quasiperiodic, influences the multifractal characteristics of the states in the Fock space. 

We find that, similar to the random disordered potential, $\Delta_t$ remains $O(1)$ in the thermal phase, and in the MBL phase, it decays with increasing system sizes following a power-law $\Delta_t\sim \mathcal{N}_F^{-(1-D_s)}$, as shown in Fig.\ref{appfig:fractal}(a). We extract the fractal dimension $D_s$ from the power-law fit to $\Delta_t(\mathcal{N}_F)$ [Fig.\ref{appfig:fractal}(b)]. The value of $D_s$ estimated from power-law fit is consistent with its value extracted from finite-size scaling as discussed below. $D_s$ obtained from finite-size scaling changes discontinuously across the critical potential strength $W_c$ from $1$ in the thermal phase to a value between $0$ and $1$ in the MBL phase, where $W_c$ is also estimated from finite-size scaling.

\begin{figure}[h!]
    \centering
    \includegraphics[width=0.5\textwidth]{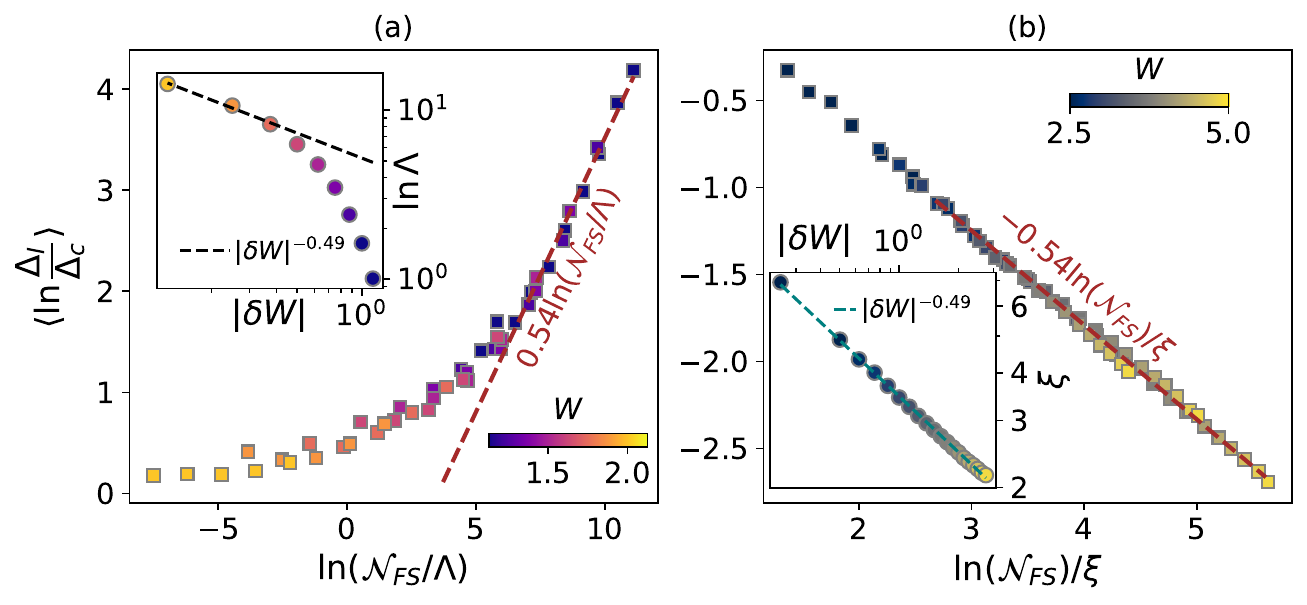}
    \caption{Scaling collapse for the typical value of the Feenberg self-energy in (a) the thermal phase and (b) the MBL phase. The FS volume scale $\Lambda$ and the length scale $\xi$ extracted from the scaling collapse are shown as a function of $|\delta W|=|W-W_c|$ in the respective insets. The critical potential strength $W_c=2.25$ is estimated from the quality of the finite-size scaling collapse (see main text).} 
    \label{fig:self_en}
\end{figure}

We further analyze the finite-size results for the self-energy using an asymmetric scaling ansatz given by~\cite{Sutradhar2022.FSP,Mace2019.multifractal,roy2021.Fock,Garc_PRL2017}
\begin{equation}\label{eq:scaling_ansatz}
\ln\left(\dfrac{\Delta_t}{\Delta_c}\right)= \begin{cases}
\mathcal{G}_{\mathrm{vol}}\left(\ln \frac{\mathcal{N}_F}{\Lambda} \right) & \mathrm{for}~~ W<W_c\\
\mathcal{G}_{\mathrm{lin}}\left(\frac{\ln\mathcal{N}_F}{\xi}\right) & \mathrm{for}~~W> W_c\\
\end{cases}
\end{equation}
where $\Delta_c$ is the typical value of the self-energy at the critical point $W=W_c$. In the thermal phase, $\mathcal{G}_{\mathrm{vol}}$ is a `volumic' scaling function associated with a Fock-space volume scale $\Lambda$ while in the MBL phase, $\mathcal{G}_{\mathrm{lin}}$ is a `linear' scaling function with a Fock-space length scale $\xi$. Fig.~\ref{fig:self_en} shows the scaling collapse of the finite size data up to $L=20$ with $W_c$ chosen as $2.25$, which gives the best quality of collapse, i.e., the error of fit to the finite size scaling forms [Eq.\eqref{eq:scaling_ansatz}] to the data on both thermal and MBL sides is minimal. This estimate is also consistent with $W_c$ obtained from the many-body level-spacing ratio(not shown). Here, the scaling functions $\mathcal{G}_{\mathrm{vol}}$ and $\mathcal{G}_{\mathrm{lin}}$ provide satisfactory collapse in the thermal and MBL phases, respectively. 

In the thermal phase, the FS volume scale $\Lambda$ obtained from finite-size scaling diverges near the critical point following a KT-like essential singularity $\Lambda\sim \mathrm{exp}\left[b/(W_c-W)^a\right]$ with $b\sim O(1)$ and  $a\simeq 0.5$ [as shown in the inset of Fig.~\ref{fig:self_en}(a)], similar to the observation for the case of random disordered potential~\cite{Sutradhar2022.FSP}. Since $\Delta_t\sim O(1)$ in the thermal phase for large enough system sizes $\mathcal{N}_F\gg \Lambda$, it can be shown~\cite{Sutradhar2022.FSP} that in the asymptotic limit of $x=\mathcal{N}_F/\Lambda\gg 1$, $\mathcal{G}_{vol}\left(x\right)\sim -(1-D_c) \ln x $, $D_c$ being the fractal dimension at the critical point $W_c$. As seen in Fig.~\ref{fig:self_en}(a), $D_c=0.46$ is found from the slope of the linear fit in the asymptotic limit ($x\gg 1$). This is in agreement with $D_c$ extracted directly from the scaling of $\Delta_t$ with system size, $\Delta_t\sim \mathcal{N}_F^{-(1-D_s)}$, shown in Fig.~\ref{appfig:fractal}(a). 

In the MBL phase, the Fock-space length scale $\xi$ obtained from finite-size scaling diverges near the critical point [see the inset of Fig.~\ref{fig:self_en}(b)] following a power law $\xi \sim |W-W_c|^{-\beta}$ with $\beta\simeq 0.5$ while approaching the critical point from the MBL side. This again is very similar to what we see in the random disorder case~\cite{Sutradhar2022.FSP}.
In the limit of $x=(\ln \mathcal{N}_F)/\xi\gg 1$ the scaling function $\mathcal{G}_{lin}(x)$ asymptotically approaches~\cite{Sutradhar2022.FSP}  $\sim -(1-D_c)x$, as seen in Fig.\ref{fig:self_en}(b). Again, $D_c=0.46$ is consistent with that obtained from asymptotic scaling in the thermal side, discussed above, as well as with $D_c$ estimated in Fig.\ref{appfig:fractal}. The scaling form $\mathcal{G}_{lin}$ along with the power-law $\Delta_t\sim \mathcal{N}_F^{-(1-D_s)}$ for $W>W_c$ implies $\xi=(1-D_c)/(D_c-D_s)$, and thus a diverging $\xi$ as $D_s\to D_c$ for $W\to W_c^+$. 

In summary, we find that the differences between random and quasiperiodic disorder in real space do not give rise to any qualitative differences in fractal properties of the states across the MBL transition and the associated finite-size scaling of the typical FS Feenberg self-energy for the system sizes accessed in our study.


\subsection{Distribution of Feenberg self-energy}\label{ssec:distribution}

Since the typical Feenberg self-energy cannot distinguish between random and quasiperiodic systems in the Fock space, in this section, we look into the statistics of $\Delta_I$ in terms of its distribution over different FS sites $I$ in the middle slice and over different disorder realizations. 
An approximate self-consistent theory~\cite{Logan2019.Fock} for the self-energy predicts a log-normal distribution in the thermal phase, while deep in the MBL phase, the ratio of the self-energy to the width $\eta$ is predicted to follow a L\'evy distribution with a characteristic tail $\sim (\Delta/\eta)^{-3/2}$.
\begin{figure}
    \centering
    \includegraphics[width=0.5\textwidth]{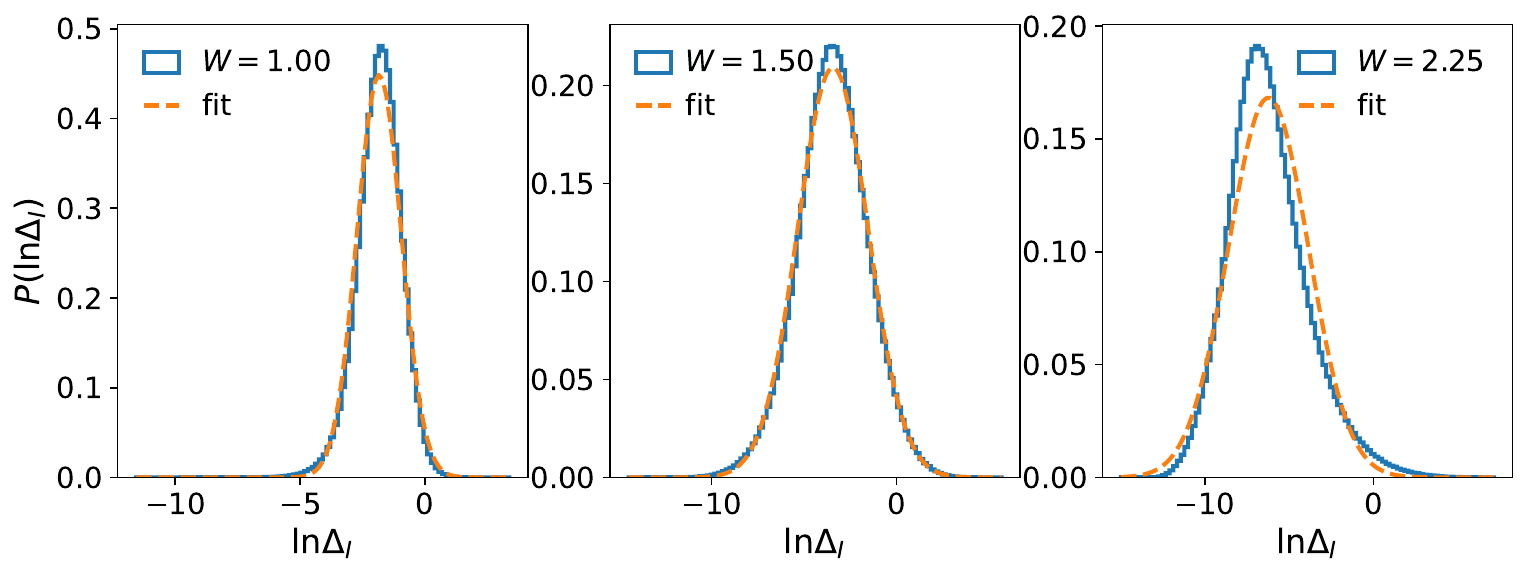}
    \caption{Distribution of the self-energy ($\mathrm{ln} \Delta_I$) for potential strengths $W=1.0$, $1.5$, $2.25$. The Gaussian fit shows the deviation from the expected log-normal functional form for $\Delta_I$.}
    \label{fig:dstr}
\end{figure}
\begin{figure}[h!]
    \centering
    \includegraphics[width=0.5\textwidth]{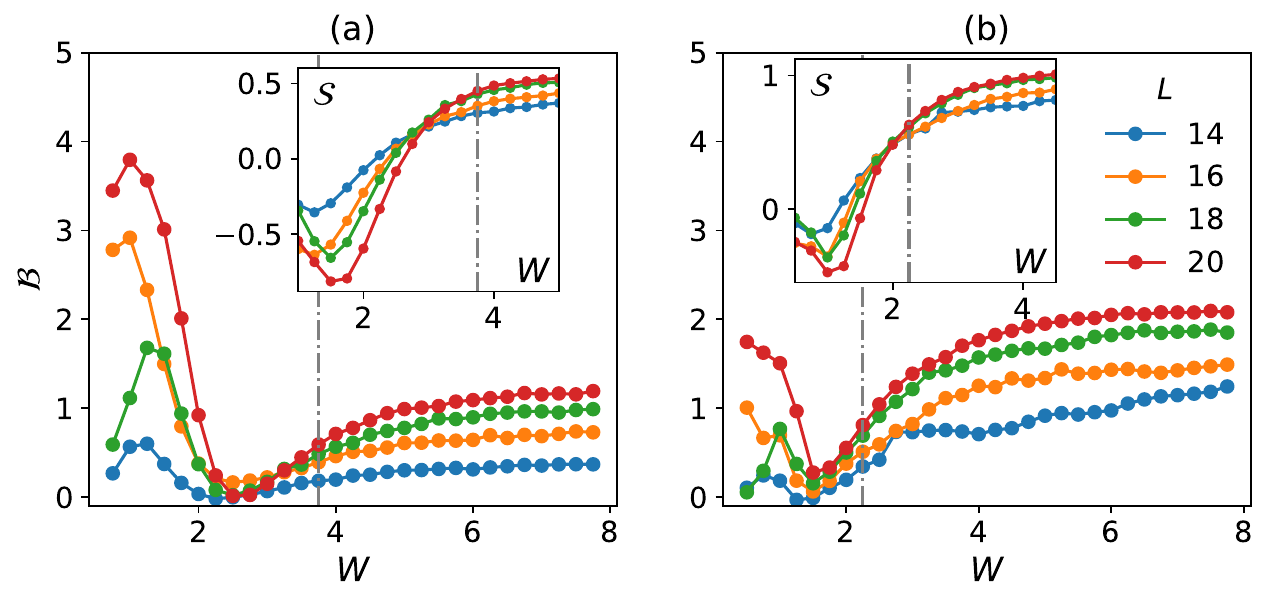}
    \caption{Binder cumulant for the distribution $P(\mathrm{ln}\Delta_I)$ as a function of the potential strength for (a) random disorder and (b) the quasiperiodic potential. Near the transition for both models, the distribution approaches a Gaussian as indicated by $\mathcal{B}\simeq 0$. The insets show the variation of the skewness $\mathcal{S}$ of the distributions for the respective cases. The skewness changes sign close to the transition. The critical potential strength $W_c$ is indicated in the plots via vertical dashed lines for reference.}
    \label{fig:binder}
\end{figure}

We find that for both potentials the distribution of self-energies $\Delta_I$ follows slightly different functional forms than those predicted by the self-consistent calculations in the thermal phase. While deep in the thermal phase the distribution of $\ln\Delta_I$ is expected to be Gaussian, we see non-Gaussian characteristics as well as a lack of symmetry in the distribution, as shown in Fig.\ref{fig:dstr} for quasiperiodic potential. Similar results are found for random system (not shown).

To quantify these further, we calculate the Binder cumulant or kurtosis for $x=\ln \Delta_I$ defined as $\mathcal{B}=\langle \delta x^4\rangle/\langle\delta x^2\rangle^2-3$ with $\delta x$ defined as $\delta x=x-\langle x\rangle$.
A Binder cumulant $\mathcal{B}=0$ indicates a perfect Gaussian distribution of the underlying variable, and the deviation from $\mathcal{B}=0$ characterizes non Gaussianity of the distribution. We also calculate the skewness defined by $\mathcal{S}=\langle \delta x^3\rangle/\langle \delta x^2\rangle^{3/2}$.
Since a Gaussian distribution is symmetric around the mean, the skewness is zero. Any non-zero skewness again indicates deviation from Gaussianity.

As shown in Fig.~\ref{fig:binder} (a), the Binder cumulant of the distribution $P(\ln\Delta_I)$ for the random disordered potential approaches zero near the critical point for the thermal-MBL transition. However, the potential strength at which the cumulant becomes $0$ is somewhat below the typical estimate of the critical disorder $W_c\simeq 3.5-4.0$ \cite{Luitz.2015,Sutradhar2022.FSP,Sierant2022} for the random case considered here. The skewness of the distribution [inset of Fig.\ref{fig:binder}] also changes sign close to the transition indicating that the distribution $P(\ln\Delta_I)$ changes from being left skewed to right skewed as the disorder strength is changed. However, curves corresponding to different system sizes cross at a value of $W$ where the skewness $\mathcal{S}$ is non-zero but small. This together with the Binder cumulant implies the distribution becomes Gaussian close to the transition. But the values of $W$ at which this Gaussianity emerges seem to lie within the thermal side, somewhat below the estimated range of critical disorder for the putative MBL transition~\cite{Luitz.2015,Sutradhar2022.FSP,Sierant2022}.

Similar features are seen for the quasiperiodic potential in Fig.~\ref{fig:binder}(b). Here, although a decrease in the Binder cumulant is observed close to the thermal-MBl transition, the Binder cumulant never actually reaches zero, especially for the largest system size ($L=20$), unlike in the case of the random potential. Accordingly the system size variation of skewness shows a crossing near the thermal-MBL transition. But, the value of skewness at the crossing point is substantially larger than that for the random case. Nevertheless, the overall variation of both quantities appears qualitatively similar to the case of the random disordered potential, namely the distribution $P(\ln\Delta_I)$ approaches a Gaussian [or, $P(\Delta_I)$ a log-normal] distribution in the thermal phase before the MBL transition, whereas non-Gaussianity increases with system size deep in the thermal and MBL phase. Thus, the distribution $P(\Delta_I)$ never truly becomes log-normal even deep inside the thermal phase. This is in contrast to tight-binding model on regular lattices with uncorrelated disorder, where the log-normal distribution is ubiquitous, both in the localized and delocalized phases \cite{Schubert2010}, and naturally arises in the non-linear $\sigma$ model description of such systems \cite{Mirlin1996,Mirlin2000}. On the FS lattice, however, the disorder is highly correlated \cite{Altland2017,Logan2019.Fock,Roy2020,Ghosh2019.correlated}, and there is no reason for such descriptions for an uncorrelated disorder to be applicable, as clearly suggested by our results.


\subsection{Fock space localization length from off-diagonal elements of the Green's function}\label{ssec:FS_llength}
Here we discuss the scaling of the non-local elements of the Green's function. We are again interested in the middle slice of the FS lattice. As already mentioned, we consider two types of off-diagonal elements of the Green's function: (i) a typical $G_{IJ}=\langle I|\mathbf{G}|J\rangle$, where both $I$ and $J$ are FS sites belonging to the middle slice, and correspond to generic states, and (ii) a non-generic $G_{1I}= \langle 1|\mathbf{G}|I\rangle$, where $I$ belongs to the middle slice and $|1\rangle$ is an atypical FS site having the minimal connectivity in the FS; $|1\rangle$ corresponds to a single domain wall state where all the sites on the left are occupied and the sites on the right are empty. Thus, this is a non-generic state for energy $E=0$ in the middle of the many-body spectrum. 
The rationale behind the choice of only two types of off-diagonal elements of FS propagator are (a) computational convenience since the recursive Green's function method is most efficient when only a few types of elements are targeted, and (b) the fact the above two types of elements represent two extreme cases, namely propagator involving two typical sites and that between an atypical and a typical one. 

To quantify $G_{IJ}$, we use the hopping distance $r_{IJ}$ as a notion of distance in FS.
The hopping distance $r_{IJ}$ is given by the minimum number of hops that connect FS sites $I$ and $J$. Due to the construction of FS (Fig.~\ref{fig:FS-lattice}), the hopping distance from $1$ to site $I$ is the same for all sites $I$ belonging to the same slice. Thus, $r_{1I}\propto L^2$ for all sites belonging to the middle slice of FS. On the other hand, the hopping distance $r_{IJ}$ between typical FS sites $I$ and $J$ can vary from $2$ to $r_{max}$ where $I$ and $J$ both belong to the middle slice. The maximum hopping distance $r_{max}$ depends on the system size $L$ and the number of particles, as discussed in  Appendix \ref{app:G_MM}.
\begin{figure}[h!]
    \centering
    \includegraphics[width=0.5\textwidth]{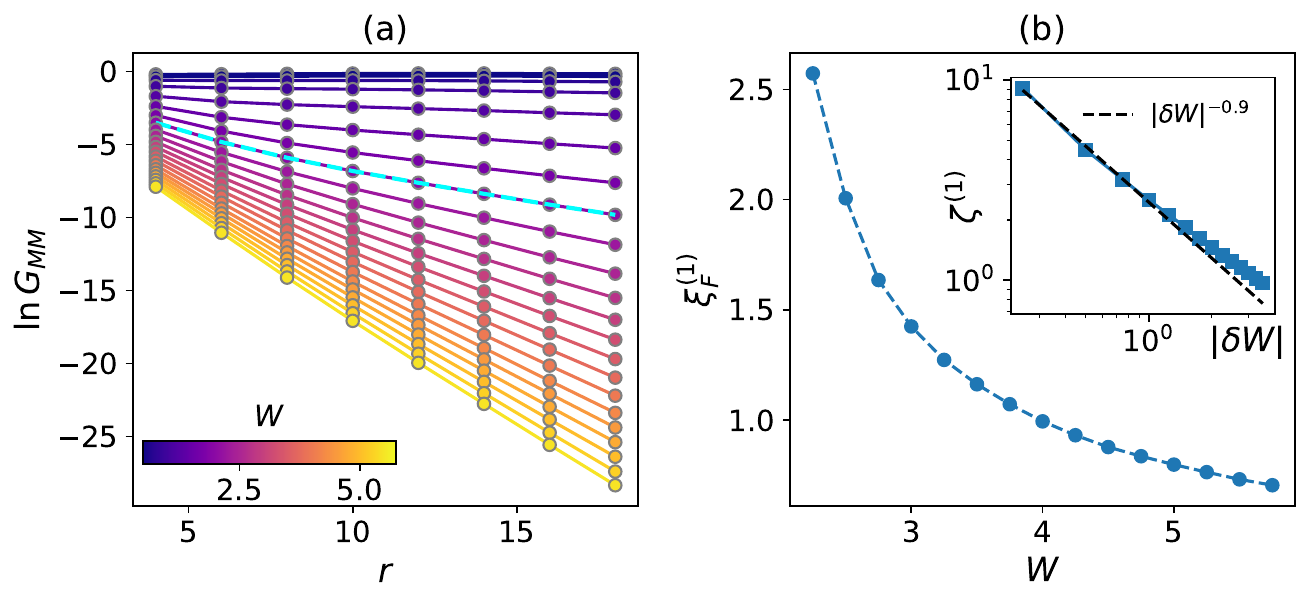}
    \caption{(a) Variation of the typical value of the off-diagonal elements $G_{IJ}$ of the Green's function as a function of the hopping distance $r$ for different values of the potential strength $W$ for system size $L=20$. (b) Scaling of the Fock-space localization length $\xi_F^{(1)}$ as a function of the potential strength for $W>W_c$. The inset shows the correlation length $\zeta^{(1)}$ that varies as $\zeta^{(1)}\sim |W-W_c|^{-\nu}$ with $\nu\sim 0.9$.}
    \label{fig:xi_GMM}
\end{figure}

\textit{Typical FS localization length.}--- We calculate the typical value of off-diagonal elements $G_{IJ}$ for $I$ and $J$ separated by hopping distance $r_{IJ}=r$ as $\ln G_{MM}(r)=\langle \ln |G_{IJ}(r_{IJ})|\rangle_{r_{IJ}=r;\phi}$ where $\langle\cdots\rangle_{r_{IJ}=r;\phi}$ denotes averaging over different pairs of $I$ and $J$ with the same hopping distance $r$ and over different realizations of the offset angle $\phi$. Fig.~\ref{fig:xi_GMM}(a) shows the variation of the typical value $G_{MM}(r)$ as a function of $r$. In the thermal phase ($W<W_c$), $\ln G_{MM}$ is almost independent of $r$, while, in contrast, $G_{MM}$ decreases exponentially with $r$ in the MBL phase.
\begin{figure}[h!]
    \centering
    \includegraphics[width=0.5\textwidth]{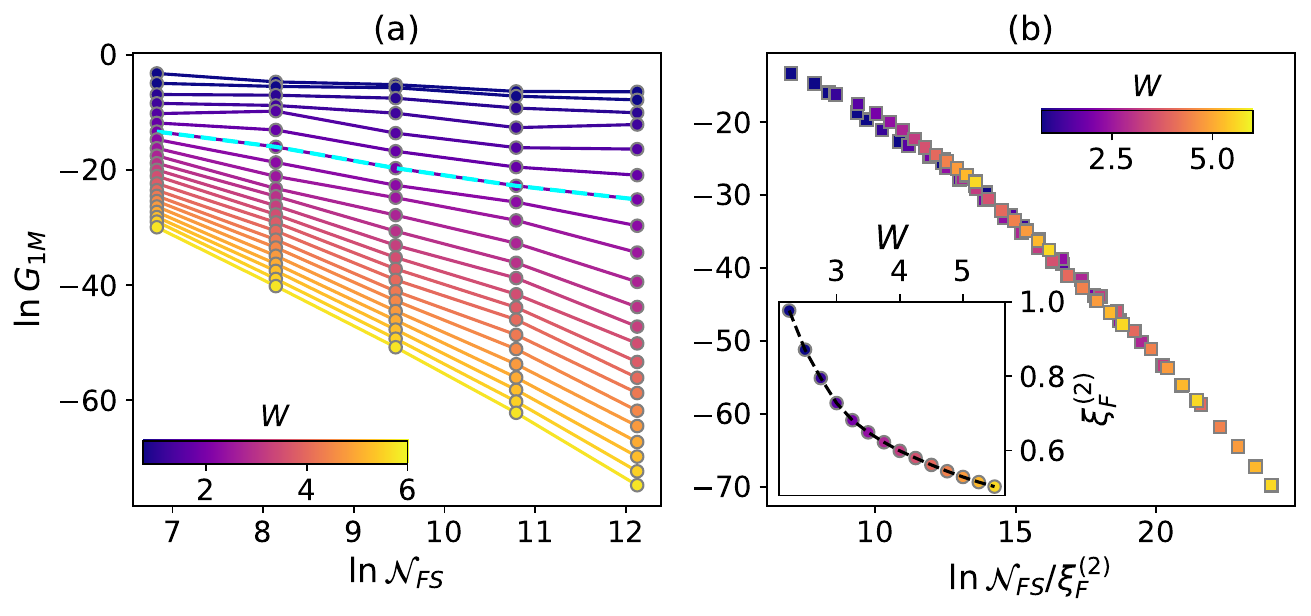}
    \caption{Variation of the typical value of off-diagonal elements $G_{1I}$ of the Green's function as a function of Fock space dimension $\mathcal{N}_{F}$ for different values of the potential $W$. (b) Scaling collapse of the typical value $\ln G_{1M}$ assuming a functional form $F(\ln \mathcal{N}_{F}/\xi_F^{(2)})$ for $W>W_c$. The inset shows the variation of the localization length $\xi_F^{(2)}$ as a function of the potential strength $W$.}
    \label{fig:xi_G1M}
\end{figure}

In the MBL phase, the eigenstates have non-zero amplitude on $\mathcal{N}_F^D$ number of sites that diverge in the thermodynamic limit. Here $D$ ($0<D<1)$ is a fractal dimension that can be extracted, e.g., from inverse participation ratio (IPR) \cite{Mace2019.multifractal,roy2021.Fock}, and is expected to be close to the spectral fractal dimension $D_s$ \cite{Altshuler2016a,Nroy2022.NEE} discussed in Sec.\ref{ssec:self_en}. However, the fraction ($\sim \mathcal{N}_F^{-(1-D)}$) of such FS sites goes to zero in the thermodynamic limit indicating the multifractality of these states. Nevertheless, we find that the typical off-diagonal propagator $G_{MM}$, which captures the correlations of amplitudes at different sites, decreases exponentially with the hopping distance $r$ in the MBL phase. Thus, we can define a Fock space localization length $\xi_F^{(1)}(W,L)$ associated with the exponential decay of $G_{MM}(r)$ through
\begin{equation}\nonumber
 G_{MM}(r)\sim \exp{\left(-\frac{r}{\xi_F^{(1)}}\right)}
\end{equation}
We call $\xi_F^{(1)}$ a typical FS localization length since it involves correlations of amplitudes between two generic sites.

Fig.~\ref{fig:xi_GMM}(b) shows the variation of $\xi_F^{(1)}$ as a function of $W$ for the largest system size, $L=20$. We find $\xi_F^{(1)}$ to have little system size dependence (see Appendix~\ref{app:xi_f}). The localization length increases as one approaches the critical point $W_c$ from the MBL side, but seems to remain finite at the critical point with $\xi_{F,c}^{(1)}=\xi_{F}^{(1)}|_{W=W_c}\simeq 2.5$. This behaviour is similar to the FS localization length extracted from a somewhat different correlation function involving the square of the modulus of the FS amplitudes in Ref.\cite{roy2021.Fock}. The length scale $\xi_F^{(1)}$, which remains finite at the MBL transition, also mimics the length scale associated with either the local support or the associated decay of matrix elements, of the LIOMs or the $l$-bits \cite{LocalSerbyn.2013,PhenomenologyHuse.2014,ConstructingChandran.2015}. Furthermore, the value of $\xi_{F,c}^{(1)}$ is quite close to the critical value $\sim 2/\ln(2)$ expected for the LIOM length scale at the MBL transition \cite{deRoeck2017,Potirniche2019,Chandran2022.crossover}, although the direct connection between the $l$-bit length scale and the FS length $\xi_F^{(1)}$ is not clear at this stage. Nevertheless, the similarity motivates us to define another length scale $\zeta^{(1)}=[1/\xi_F^{(1)}-1/\xi_{F,c}^{(1)}]^{-1}$, which diverges at the transition. We find $\zeta^{(1)}\sim |W-W_c^+|^{-\nu}$ with $\nu=0.9$, as shown in the inset of Fig.\ref{fig:xi_GMM}(b). Following Ref.\cite{Chandran2022.crossover}, $\nu\approx 1$ suggests a similarity of $\zeta^{(1)}$ with one of the length scales associated with the probability of finding resonance between two eigenstates which differ by rearrangements of $r$ local $l$ bits.

\textit{Atypical FS localization length.}--- We now discuss the localization length that can be extracted from the non-generic off-diagonal propagator $G_{1I}$ where $I$ belongs to the the middle of FS graph (Fig.\ref{fig:FS-lattice}).
Fig.~\ref{fig:xi_G1M}(a) shows the variation of the typical value of $G_{1I}$ with system sizes. The typical value $G_{1M}$ for $G_{1I}$ is calculated as $\ln G_{1M}=\left\langle \ln G_{1I}\right\rangle_{I;\phi}$ where $\langle~.~\rangle_{I;\phi}$ denotes averaging over all FS sites belonging to the middle slice and different realizations of offset angle $\phi$. Since all sites $I$ belong to the middle slice, the hopping distances of these sites $I$ from the apex site $1$ are the same. Thus, we cannot extract a length scale from the variation of the $G_{1M}$ with hopping distance, as in the case of $G_{MM}$ discussed above. 

However, a length scale can be extracted from the system size dependence of $G_{1M}$. As shown in Fig.\ref{fig:xi_G1M}(a), in the thermal phase, the typical value $G_{1M}$ is almost independent of the system size. On the other hand, in the MBL phase, the typical value decreases with the Hilbert space dimension $\mathcal{N}_F$. We assume a scaling form $\ln G_{1M}\sim f(\ln \mathcal{N}_F/\xi_F^{(2)})$ to collapse $\ln G_{1M}$ for different $W$ and different system sizes. Here $\xi_F^{(2)}$ denotes an atypical FS localization length. The good quality of the scaling collapse shown in Fig.~\ref{fig:xi_G1M}(b) indicates that $\ln G_{1M}$ indeed is function of $\ln \mathcal{N}_F/\xi_F^{(2)}$. The extracted localization length is shown in the inset of Fig.~\ref{fig:xi_G1M}(b). Very similar to $\xi_F^{(1)}$, $\xi_F^{(2)}$ also decreases with increasing potential strength in the MBL phase ($W>W_c$), approaching a finite value $\xi_{F,c}^{(2)}$ at $W_c$. However, $\xi_{F,c}^{(2)}$ substantially smaller than $\xi_{F,c}^{(1)}$. Again, a length scale $\zeta^{(2)}$ defined by $\zeta^{(2)}=[1/\xi_F^{(2)}-1/\xi_{F,c}^{(2)}]^{-1}$ diverges near the critical point $W_c$ following a power law $|W-W_c|^{-\nu}$ with $\nu=0.9$ (not shown). Therefore, the atypical length scales $\xi_{F}^{(2)}$ and $\zeta^{(2)}$ exhibit similar critical properties as $\xi_{F}^{(1)}$ and $\zeta^{(1)}$, respectively, and might also be related to length scales associated with LIOMs. However, the exponent $\nu\approx 1$ associated with divergence of the length scales $\zeta^{(1)},~\zeta^{(2)}$ at $W_c^+$ appear to be discernibly different from the exponent $\beta\approx 0.5$ for the length scale $\xi$ extracted from the finite-size scaling of the typical value of Feenberg self-energy in Sec.\ref{ssec:self_en}. We also note that both the exponents, particularly $\beta$, associated with diverging FS length scales violate the Harris-Luck bound~\cite{Luck1993}, which requires $\beta,\nu\geq 1/d$ for $d$-dimensional quasiperiodic systems. This is believed to be a common artifact of small system sizes accessed in all the \emph{ab initio} numerical studies of MBL systems, e.g., via ED. The behaviours of the FS localization lengths for quasiperiodic disorder discussed here are found to be very similar to that for the random disorder case \cite{Greens_new}.
\begin{figure}[h!]
    \centering
    \includegraphics[width=0.5\textwidth]{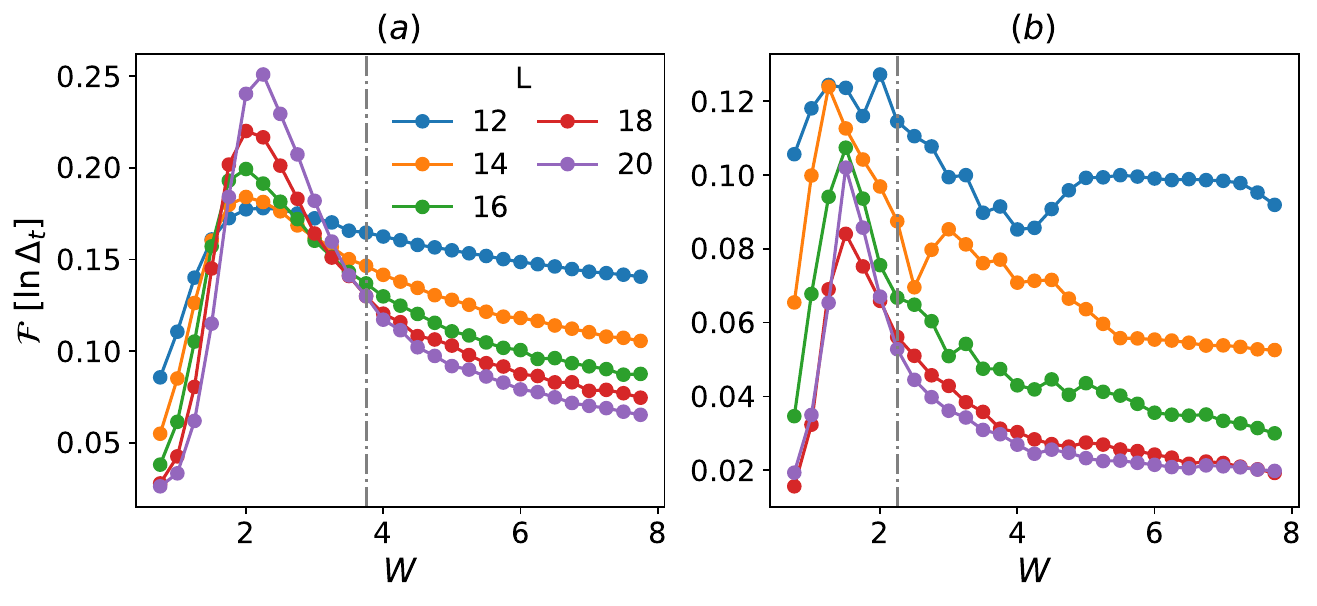}
    \caption{Sample to sample fluctuations in the typical value of the Feenberg self-energy $\ln \Delta_t$ for (a) random disorder and (b) the quasiperiodic potential. The peak in the fluctuations increases with system size and becomes sharper near the thermal-MBL transition in the random disorder case. In contrast, in the case of the quasiperiodic potential, while the fluctuation peaks near the transition, the peak height decreases with system size instead of increasing. The critical potential strength is indicated by the vertical dashed line in the plots for reference.}
    \label{fig:fluct_selfen}
\end{figure}

\subsection{Sample-sample fluctuations of FS propagators for quasiperiodic and random disorders}\label{ssec:fluct}

In the previous subsection, we have shown that the typical values associated with distributions of quantities related to the local and the non-local parts of FS Green's function over various FS sites and disorder realizations exhibit quite sharp signatures of MBL transition in terms of different diverging length scales, at least for the finite systems accessed in this work. These analyses, though give important insights into the structure of the Green's function in the FS, do not provide any distinctions between random and quasiperiodic disorders from the perspective of Fock space. 
However, the quasiperiodic potential is very different from the true random disorder owing to the deterministic nature of the former. Since the typical values fail to capture any difference between the two, we analyze the quantities of interest further by calculating the sample-to-sample fluctuations. In Ref.\onlinecite{Khemani2017}, the MBL transitions in the random and quasiperiodic systems have been argued to be in different universality classes due to differences in fluctuations of entanglement entropy from one realization of potential (\emph{a sample}) to other, and fluctuations within a given sample over different eigenstates.

Here, we define the sample-to-sample fluctuation of the Feenberg self-energy as $\mathcal{F}[\ln \Delta_t]=\sigma~[\ln \Delta_t^{(s)}]/\ln \Delta_t$, where $\ln\Delta_t^{(s)}=\left\langle\ln \Delta_I\right\rangle_{I}^{(s)}$ is the typical value for a given sample $s$ for the distribution of $\Delta_I$ over different sites in the middle slice (Fig.~\ref{fig:FS-lattice}), and $\sigma~[\ln\Delta_t^{(s)}]$ is the standard deviation of these typical values over different samples.

For the non-local part of the Green's function, we analyze the fluctuation in the quantity $G_{IJ}$ for $r_{IJ}=L-2$. The fluctuation is defined as $\mathcal{F}\left[\ln G_{MM}\right]=\sigma~[\ln G_{MM}^{(s)}]/\ln G_{MM}$, where $\ln G_{MM}^{(s)}=\left\langle\ln G_{IJ}\right\rangle_{I,J}^{(s)}$ is the typical value calculated for each sample $s$ by averaging over all the pairs $I,J$ with hopping distance $r_{IJ}=L-2$. Similar to self-energy, $\sigma[\ln G_{MM}^{(s)}]$ is the standard deviation of the distribution of $\ln G_{MM}^{(s)}$ over different samples. 
Here, we only consider the FS site pairs with hopping distance $L-2$. This choice is not special; any $r_{IJ}<L$ gives similar results. The statistics become poor for FS site pairs with hopping distances larger than $L-2$, as the number of such pairs becomes very small for all the system sizes accessed here [see Appendix \ref{app:G_MM}]. Fluctuations of the atypical element $G_{1M}$ give very similar results, as shown in Appendix \ref{app:G_1M}.
\begin{figure}[h!]
    \centering
    \includegraphics[width=0.5\textwidth]{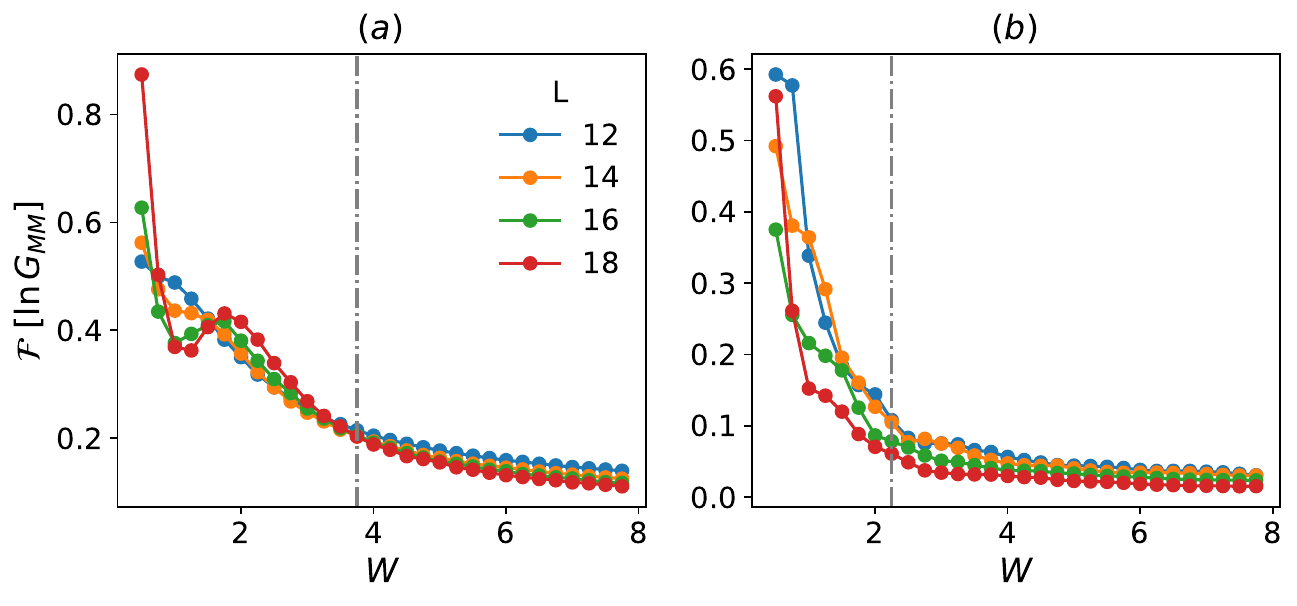}
    \caption{Sample to sample fluctuation in the typical value of off-diagonal elements $G_{IJ}$  ($\ln G_{MM}$) for the random disorder (a) and the quasiperiodic potential (b). The fluctuation changes almost monotonically as a function of potential strength $W$. The critical potential strength is indicated by the vertical dashed line in the plots for reference.}
    \label{fig:fluct_GMM}
\end{figure}

Fig.~\ref{fig:fluct_selfen} shows the fluctuation in self-energy as a function of the potential strength $W$. Deep in the thermal phase and MBL phase, the fluctuations are very small and they decrease with $L$. The fluctuation peaks near, but somewhat below, the critical point $W_c$ on the thermal side. Incidentally, for both quasiperiodic and random systems, the peak position is very close to the potential strength where the distribution of Feenberg self-energy becomes nearly Gaussian, as discussed in Sec.\ref{ssec:distribution}. In the presence of true random disorder [Fig.~\ref{fig:fluct_selfen}(a)], the fluctuation's peak height increases rapidly with increasing system size and the peak becomes sharper. On the contrary, in the case of the quasiperiodic potential [Fig.~\ref{fig:fluct_selfen}(b)], though the sample-to-sample fluctuation of the Feenberg self-energy exhibits a peak near the transition, the peak height diminishes with increasing system size. 

This increasing fluctuation in the random disorder case might be attributed to Griffiths-like rare region effects \cite{Gopalakrishnan2016,Agarwal2015,Agarwal2017}, known to occur in the presence of true random disorder. These are related to the presence of rare localized (thermal) regions in an otherwise thermal (localized) system due to finite but large regions with atypical strong (weak) disorder near the transition for $W<W_c$ ($W>W_c$). In the thermal phase, these rare regions hinder transport and strongly affect characteristic time scales related to relaxation, whereas, in the MBL phase, they act like a finite-size bath \cite{Agarwal2017,deRoeck2017}. In fact, these rare region effects are known to dominate transport over an extended range of disorder near the MBL transition, but somewhat below it, on the thermal side \cite{Agarwal2015,Agarwal2017}. Thus, in the thermal phase near $W_c$, disorder realizations with such rare regions are expected to have a very small typical value of the self-energy compared to a typical disorder realization. Therefore, the sample-to-sample fluctuations of the typical Feenberg self-energy may become very large approaching $W_c$ from the thermal phase with increasing system size in the case of true random disorder. In contrast, for the deterministic quasiperiodic potential, naively, one does not expect any rare regions to exist and no such enhancement of the fluctuations of Feenberg self-energy.


Fig.~\ref{fig:fluct_GMM} shows that the sample-to-sample fluctuations in $G_{MM}$ vary almost monotonically with increasing potential strength $W$, unlike the fluctuations in Feenberg self-energy. However, in close inspection, a weak \emph{hump} like feature seems to develop for larger systems in the thermal region near the transition. Nevertheless, there is not much qualitative difference between the random disordered and quasiperiodic potentials when viewed through the lens of fluctuations in the off-diagonal elements of Green's functions. Similar variations are seen in the off-diagonal elements $G_{1I}$ as shown in Appendix \ref{app:G_1M}. Thus, the FS localization lengths $\xi_{F}^{(1)},~\xi_{F}^{(2)}$, which remain finite at the transition, seem to be more or less unaffected by the putative rare region effects, in contrast to the characteristic rate or time scale obtained from the Feenberg self-energy.

\section{Conclusion}\label{ssec:conclusion}
In this paper, we provide a characterization of the thermal and MBL phases, and the phase transitions between them, in the Fock space for a one-dimensional quasiperiodic system in terms of the many-body FS propagator or Green's function. In particular, we focus on how the difference of real-space correlations, or lack thereof, associated with quasiperiodic and random systems manifests in Fock space. To this end, we compute the diagonal and off-diagonal elements of the FS propagator at an energy in the middle of the many-body spectrum using an efficient recursive Green's function method that can access system sizes comparable to state-of-the-art exact diagonalization studies \cite{polfed,Luitz.2018}. 
We show that the Feenberg self-energy extracted from the diagonal part of the Fock space propagator follows different finite-size scaling forms in the thermal and the MBL phases. In the MBL phase for the quasiperiodic system, the self-energy captures characteristic features of the multifractality of the MBL eigenstates, previously seen for the random case \cite{Sutradhar2022.FSP}. The typical off-diagonal elements of the FS propagator also vary quite differently in the two phases. We extract typical and atypical FS localization lengths from the off-diagonal propagator connecting different parts of the FS graph and show that these length scales obey similar critical properties. In particular, the FS localization lengths increase approaching the transition from the MBL phase, but remain finite at the transition, closely mimicking the real-space length scales associated with $l$-bits that appear in the effective description of the MBL phase \cite{Serbyn.2015,PhenomenologyHuse.2014,ConstructingChandran.2015}. However, these features are very similar to the observations in the case of random disorder~\cite {Sutradhar2022.FSP,Nroy2022.NEE}, and, thus, the typical diagonal and off-diagonal FS propagators cannot distinguish quasiperiodic and random systems in Fock space. However, we show that the sample-to-sample fluctuations of the typical Feenberg self-energy vary differently as a function of potential strength and system size for the quasiperiodic and random systems.

The fluctuation in self-energy is found to be small deep in the thermal phase and MBL phases, whereas the fluctuation exhibits a peak on the thermal side near the transition. For the random system, the peak becomes higher and sharper with increasing system size, although the fluctuation decreases with system size deep in the phases. In contrast, the peak and the overall fluctuation for the entire range of potential strength monotonically decreases with system size for the quasiperiodic system.
We discuss the plausible connection of the different system size dependence of the inter-sample fluctuation of self-energy for random and quasiperiodic systems with the presence and absence of rare regions, respectively. We also show that the sample-to-sample fluctuation in the off-diagonal elements of the FS propagator associated with the FS localization lengths does not capture any difference for the two different potentials.

In recent years, the stability of the MBL phase, even in one dimension, has been questioned and intensely debated \cite{vsuntajs2020quantum,Polkovnikov2021,Sels2022}. Evidence of long-range many-body resonances \cite{Morningstar2022}, that almost certainly push the MBL transition to a much higher value of critical disorder or destroy the MBL phase altogether \cite{Polkovnikov2021,Sels2022}, has been found. In the latter case, the sharp signatures of the MBL transition found here, or in many other previous works using, e.g., ED \cite{Luitz.2015}, for finite-systems should be understood as a finite-size crossover that would go away when the system size becomes large enough to accommodate the long-range resonances \cite{Chandran2022.crossover}. The MBL phase and transition in this case should be dubbed as \emph{finite-size MBL} phenomena. Nevertheless, the finite-size MBL phenomenology will still remain very much relevant for experimentally accessible system sizes \cite{schreiber2015observation}, as the destabilization of the MBL by the long-range resonances will presumably require humongous system sizes, especially for strong enough disorder. 

In the scenario where the resonances only shift the critical disorder to a higher value, we can broadly classify the mechanisms for the thermal-MBL transition into two further scenarios, though not mutually exclusive. In one of them, the rare regions of weak disorder in an otherwise localized system occur due to the randomness of the disorder potential. These rare regions proliferate~\cite{Morningstar2019,Goremykina2019,Dumitrescu2019} with decreasing disorder strength finally destabilizing the MBL phase through the `quantum avalanche' procedure \cite{deRoeck2017} in random systems. These rare regions are absent in the presence of deterministic potentials like the quasiperiodic potentials. Thus, the  avalanche instability is not expected to occur in the quasiperiodic systems, unlike the random case. However, in interacting quasiperiodic systems, Hartree-Fock shifts in the many-body states or in the initial condition for non-equilibrium time evolution, can still lead to effective rare regions \cite{Agrawal2022}. Numerical ED studies \cite{Khemani2017}, as well as  phenomenological real-space RG studies \cite{Zhang2018,Zhang2019}, have shown evidence of different universality classes for MBL transition in quasiperiodic systems compared to the random ones. Nevertheless, the phenomenological RSRG studies point to different critical properties for quasiperiodic MBL transition than the random case \cite{Zhang2018,Zhang2019}. 

In the alternative picture~\cite{Chandran2022.crossover} for the MBL transition, a phenomenological random resonance model has been developed without explicitly assuming the presence of rare regions. Here many-body resonances have been argued to destabilize the MBL phase for large systems till much stronger disorder, e.g., $W\gg 3.7$. For system sizes smaller than the characteristic length scale~\cite{Chandran2022.crossover} of the resonances, like the system sizes accessed via ED and other related methods, only finite-size crossover between the thermal and MBL phases are observed. Since the ED studies are limited to relatively small finite-size systems and the real-space RG methods \cite{Potter2015,VoskAltman,Morningstar2019,Goremykina2019,Dumitrescu2019} relies on phenomenological rules that cannot be derived directly from the microscopic models, the questions on the mechanism of MBL transitions in both random and quasiperiodic systems and essential differences between MBL phenomena in the two types of systems remain elusive. Our work, which provides a Fock space approach to the quasiperiodic MBL and its similarities and differences from MBL phenomena in random systems, may be useful to build better insights into the outstanding unresolved questions through a perspective complimentary to the real-space approaches.

\section{Acknowldegements}
We acknowledge useful discussions with Nilanjan Roy, Sthitadhi Roy and David Logan during collaboration on related topics.
SM and SB thank the QuST project of the DST, Govt. of India for support. S. B. acknowledges support from SERB (Grant No. CRG/2022/001062), DST, India.

\appendix

\section{Dependence of the off-diagonal element $G_{IJ}$ on hopping distances}\label{app:G_MM}
\begin{figure}[h!]
    \centering
    \includegraphics[width=0.35\textwidth]{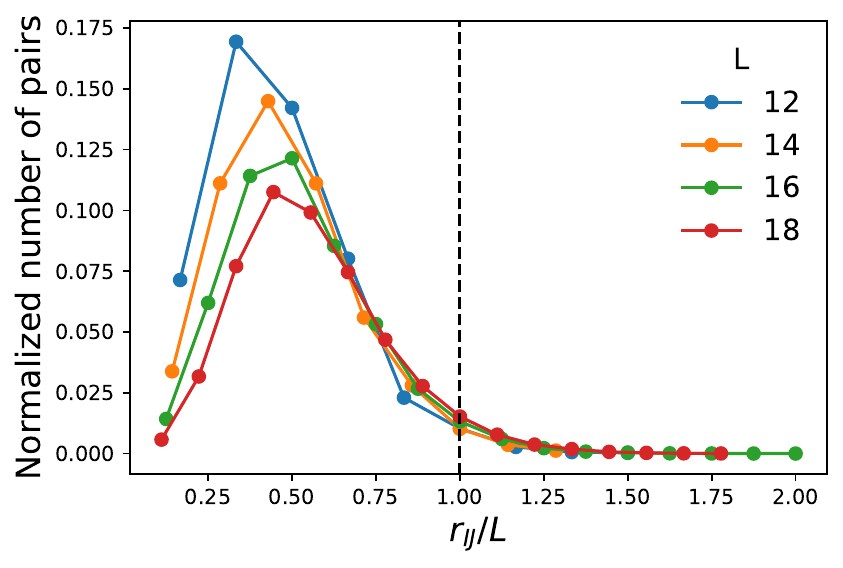}
    \caption{Normalized density of the number of FS lattice pairs as a function of the hopping distance between them. The number of FS lattice pairs in the middle slice having a hopping distance larger than $L$ is negligibly small for all system sizes. The x-axis is rescaled by the system size for a better comparison between different system sizes.}
    \label{fig:hopp_pair}
\end{figure}

Here we discuss the distribution of the number of pairs of Fock space sites $I,J$ as a function of the hopping distance $r_{IJ}$. For any system size $L$ and Fock space dimension $\mathcal{N}_F=\binom{L}{L/2}$, the number of pairs peaks at some hopping distance after which it decreases rapidly. The largest possible hopping distance $r_{max}$ for sites belonging to the middle slice depends both on system size and the number of particles. Fig.~\ref{fig:hopp_pair} shows the normalized distribution of these hopping distances. The number of pairs with $r_{IJ}\geq L$ is negligibly small for all system sizes.

\begin{figure}[h!]
    \centering
    \includegraphics[width=0.5\textwidth]{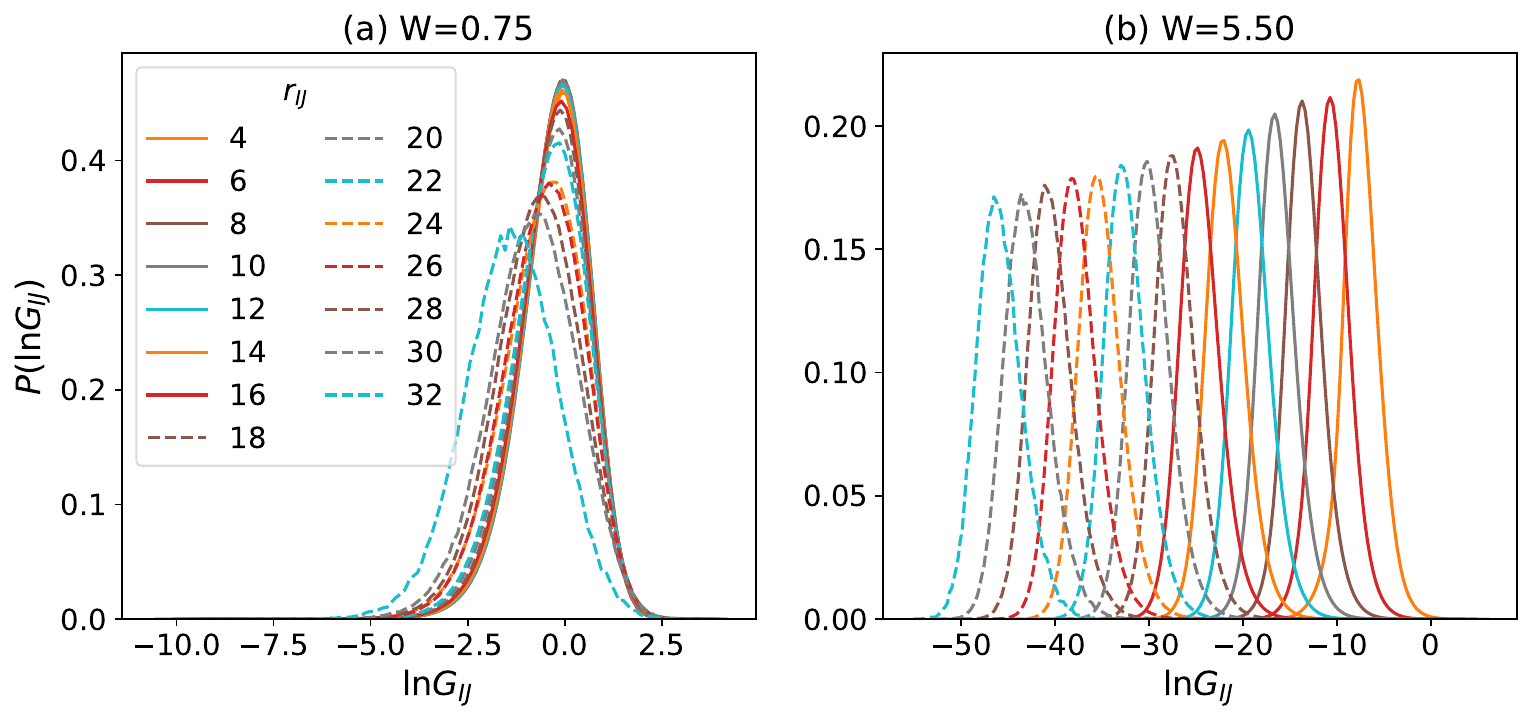}
    \caption{Distribution of $\ln G_{IJ}$ over different pairs of $I,J$ connected by hopping distances $r_{IJ}$ at system size $L=18$. The solid lines are for hopping distances $r_{IJ}<L$, while the dashed lines are for hopping distances $r_{IJ}\geq L$. Panel (a) shows the distributions deep in the thermal phase ($W=0.75$), while panel (b) shows the same deep in the MBL phase ($W=5.5$).}
    \label{fig:distr_GIJ}
\end{figure}

Fig.~\ref{fig:distr_GIJ}  shows the distribution of $\ln G_{IJ}$ over different pairs $I, J$ for different choices of $r_{IJ}$. Both in the thermal phase (a) and the MBL phase (b), the distribution of $\ln G_{IJ}$ looks very similar for different choices of $r_{IJ}$. Deep in the thermal phase, the distribution is essentially independent of the hopping distance $r_{IJ} < L$. For $r_{IJ} \geq L$, the distribution deviates from that for smaller hopping distances and picks up a dependence on $r_{IJ}$. Further, the distribution for hopping distances close to $r_{max}$ becomes ill-defined for some system sizes (not shown here). For these reasons, we analyze the behavior of typical value $\ln G_{MM}$ only up to hopping distances $L$.

\section{Fluctuation in $\ln G_{1M}$}\label{app:G_1M}
\begin{figure}
    \centering
    \includegraphics[width=0.5\textwidth]{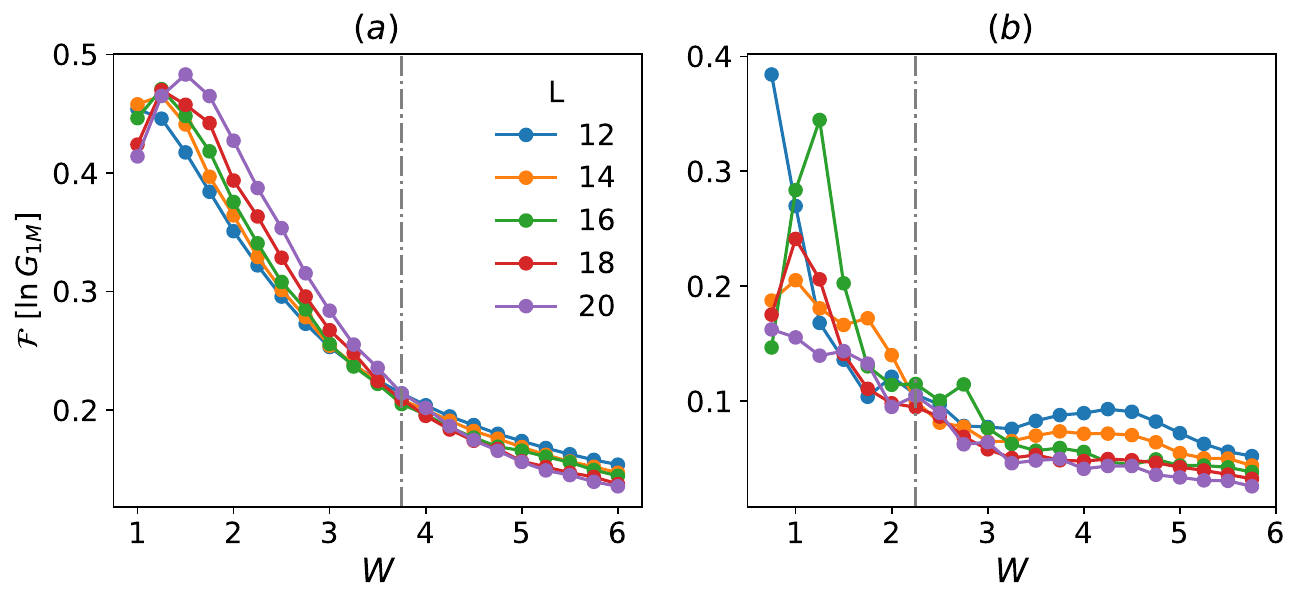}
    \caption{Sample-sample fluctuation in the typical value of off-diagonal element $G_{1I}$ ($\ln G_{1M}$) for (a) the random disorder and (b) the quasiperiodic potential. The fluctuation changes monotonically as a function of potential strength $W$.}
    \label{fig:my_label}
\end{figure}

We have discussed the sample-to-sample fluctuations of the typical value of the generic off-diagonal element $G_{IJ}$ between two FS sites $I,J$ in Sec.\ref{ssec:fluct}. Here, in Fig.\ref{fig:my_label}, we show similar results for the element $G_{1I}$ between the atypical apex site $1$ of the FS graph (Fig.~\ref{fig:FS-lattice}) and a site $I$ in the middle slice.
\section{System size dependence of Fock space localization length}\label{app:xi_f}
Here, we show the variation of the typical Fock space localization length $\xi_F^{(1)}$ with potential strength $W$ ($>W_c$) for two different system sizes $L=18$, $20$. We extract the $\xi_F^{(1)}$ from $G_{MM}(r)$ using $G_{MM}(r)\sim \mathrm{exp}\left(-r/\xi_F^{(1)}\right)$. We also show the atypical FS length $\xi_F^{(2)}$ extracted from $G_{1M}$ as discussed in Sec.~\ref{ssec:FS_llength}. The atypical FS localization length decreases with increasing potential strength in the same qualitative manner as $\xi_F^{(1)}$ but has significantly smaller values than $\xi_F^{(1)}$.
\begin{figure}
    \centering
    \includegraphics[width=0.25\textwidth]{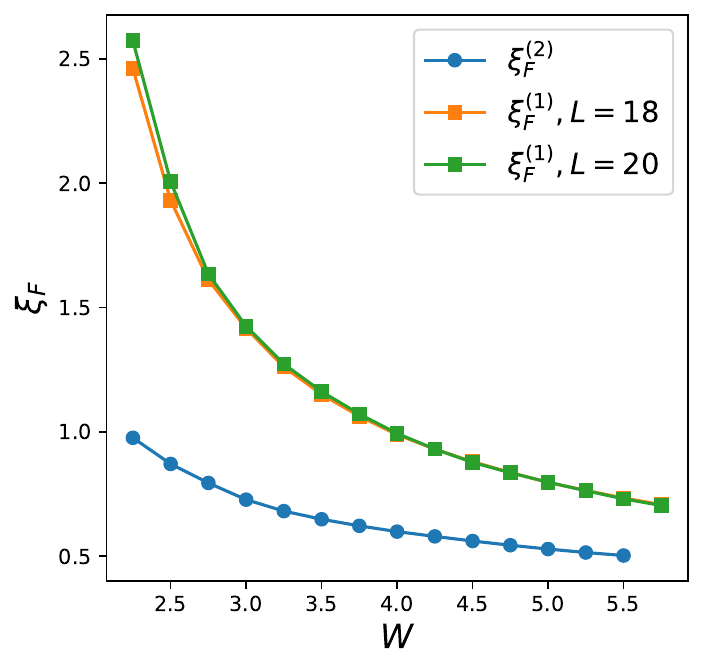}
    \caption{Variation of the FS localization length $\xi_F^{(1)}$ extracted from $G_{MM}$ for different system sizes $L=18$, $20$. It shows similar qualitative behavior as the other FS localization length scale $\xi_F^{(2)}$ extracted from $G_{1M}$.}
    \label{fig:xi_f_compare}
\end{figure}
\bibliography{reference}
\end{document}